\documentclass[onecolumn]{IEEEtran}
\usepackage{calc}
\usepackage[T1]{fontenc}

\usepackage[overload]{empheq}
\usepackage{cleveref}
\usepackage{multirow}
\usepackage{cite}
\usepackage{graphicx,subfigure}
\usepackage{psfrag}
\usepackage{amsmath,amssymb}
\usepackage{color}
\usepackage{tikz}
\usepackage{cleveref}
\usepackage{empheq}
\usepackage{dsfont}
\usepackage{caption}
\usepackage{algorithm}
\usepackage{algpseudocode}
\usepackage[utf8]{inputenc}
\usepackage[english]{babel}
\usepackage{setspace}
\newcommand*{\QEDA}{\hfill\ensuremath{\square}}

\DeclareMathOperator*{\defeq}{\triangleq}

\newtheorem{lemma}{Lemma}

\newcommand{\bit}{\begin{itemize}}
\newcommand{\eit}{\end{itemize}}

\newcommand{\bc}{\begin{center}}
\newcommand{\ec}{\end{center}}

\newcommand{\ba}{\begin{array}}
\newcommand{\ea}{\end{array}}

\newcommand{\beq}{\begin{equation}}
\newcommand{\eeq}{\end{equation}}

\newcommand{\beqn}{\begin{equation*}}
\newcommand{\eeqn}{\end{equation*}}

\newcommand{\bean}{\begin{eqnarray*}}
\newcommand{\eean}{\end{eqnarray*}}
\newcommand{\bea}{\begin{eqnarray}}
\newcommand{\eea}{\end{eqnarray}}

\makeatletter
\def\blfootnote{\gdef\@thefnmark{}\@footnotetext}
\def\BState{\State\hskip-\ALG@thistlm}
\makeatother

\begin{document}
\sloppy

\title{ITENE: Intrinsic Transfer Entropy Neural Estimator}
\author{ Jingjing Zhang, Osvaldo Simeone, Zoran Cvetkovic, Eugenio Abela, and Mark Richardson} 
\maketitle

\thispagestyle{empty}







\begin{abstract}
Quantifying the directionality of information flow is instrumental in understanding, and possibly controlling, \blfootnote{J. Zhang, O. Simeone, and Z. Cvetkovic are with the Department of Engineering at King's College London, UK (emails: jingjing.1.zhang@kcl.ac.uk, osvaldo.simeone@kcl.ac.uk, zoran.cvetkovic@kcl.ac.uk). E. Abela and M. Richardson are with the Department of Basic and Clinical Neuroscience at King's College London, UK (emails: eugenio.abela@kcl.ac.uk, mark.richardson@kcl.ac.uk). J. Zhang and O. Simeone have received funding from the European Research Council (ERC) under the European Union's Horizon 2020 Research and Innovation Programme (Grant Agreement No. 725731). J. Zhang has also been supported by a King's Together award. Code can be found at https://github.com/kclip/ITENE.} the operation of many complex systems, such as transportation, social, neural, or gene-regulatory networks. The standard Transfer Entropy (TE) metric follows Granger's causality principle by measuring the Mutual Information (MI) between the past states of a source signal $X$ and the future state of a target signal $Y$ while conditioning on past states of $Y$. Hence, the TE quantifies the improvement, as measured by the log-loss, in the prediction of the target sequence $Y$ that can be accrued when, in addition to the past of $Y$, one also has available past samples from $X$. However, by conditioning on the past of $Y$, the TE also measures information that can be synergistically extracted by observing both the past of $X$ and $Y$, and not solely the past of $X$. Building on a private key agreement formulation, the Intrinsic TE (ITE) aims to discount such synergistic information to quantify the degree to which $X$ is \emph{individually} predictive of $Y$, independent of $Y$'s past. In this paper, an estimator of the ITE is proposed that is inspired by the recently proposed Mutual Information Neural Estimation (MINE). The estimator is based on variational bound on the KL divergence, two-sample neural network classifiers, and the pathwise estimator of Monte Carlo gradients.
\end{abstract}

\begin{IEEEkeywords}
Transfer entropy, neural networks, machine learning, intrinsic transfer entropy. 
\end{IEEEkeywords}

\section{Introduction}
		
\subsection{Context and Key Definitions}		
Quantifying the causal flow of information between different components of a system is an important task for many natural and engineered systems, such as neural, genetic, transportation and social networks. A well-established metric that has been widely applied to this problem is the information-theoretic measure of Transfer Entropy (TE) \cite{T:00,Vicente2011}. To define it mathematically, consider two jointly stationary random processes $\{X_t, Y_t\}$ with $t=1,2,\dots$ The TE from process $\{X_t\}$ to process $\{Y_t\}$ with memory parameters $(m,n)$ is defined as the conditional Mutual Information (MI) \cite{T:00,MNVFHMTR:13}
\begin{align} \label{def:te}
\textrm{TE}_{X\rightarrow Y}(m,n)\defeq I(X^{t-1}_{t-m}; Y_t|Y_{t-n}^{t-1}),
\end{align} 
where $X^{t-1}_{t-m}=(X_{t-m},\dots,X_{t-1})$ and $Y^{t-1}_{t-n}=(Y_{t-n},\dots,Y_{t-1})$ denote the past $m$ and $n$ samples of time sequences $\{X_t\}$ and $\{Y_t\}$. By definition \eqref{def:te}, the TE measures the MI between the past $m$ samples of process $\{X_t\}$ and the current sample $Y_t$ of process $\{Y_t\}$ when conditioning on the past $n$ samples $Y_{t-n}^{t-1}$ of the same process. Therefore, the TE quantifies the amount by which the prediction of the sample $Y_t$ can be improved, in terms of average log-loss in bits, through the knowledge of $m$ samples of process $\{X_t\}$ when the past $n$ samples of the same process $\{Y_t\}$ are also available. While not further considered in this paper, we note for reference that a related information-theoretic measure that originates from the analysis of communication channels with feedback \cite{M:90, HYT:11} is the Directed Information (DI). The DI is defined as 
\begin{align} \label{def:di}
\textrm{DI}_{X\rightarrow Y}\defeq \frac{1}{T}\sum_{t=1}^{T}I(X_{1}^{t-1};Y_t|Y_1^{t-1}),
\end{align}
where we have normalized by the number $T$ of samples to facilitate comparison with TE. For jointly Markov processes\footnote{This implies the Markov chain $Y_t-(X_{t-m}^{t-1}, Y_{t-n}^{t-1})-(X_{1}^{t-m-1}, Y_{1}^{t-n-1})$.} $\{X_t\}$, $\{Y_t\}$ with memory parameters $m$ and $n$, the TE \eqref{def:te} is an upper bound on the DI \eqref{def:di} \cite{LA:12}. 

%



The TE, and the DI, have limitations as measures of \emph{intrinsic, or exclusive}, information flow from $\{X_t\}$ to $\{Y_t\}$. This is due to the fact that conditioning on past samples of $\{Y_t\}$ does not discount the information that the past samples of $\{Y_t\}$ contain about its current sample $Y_t$: Conditioning also captures the information that can be \emph{synergistically} obtained by observing both past samples $X_{t-m}^{t-1}$ and $Y_{t-n}^{t-1}$. In fact, there may be information about $Y_t$ that can be extracted from $X_{t-m}^{t-1}$ only if this is observed jointly with $Y_{t-n}^{t-1}$. This may not be considered as part of the intrinsic information flow from $\{X_t\}$ to $\{Y_t\}$. 

\emph{Example \cite{RBBJ:18}:} Assume that the variables are binary, and that the joint distribution of the variables $(X_{t-1}, Y_{t-1}, Y_t)$ is given as $p(0,0,0)=p(0,1,1)=p(1,0, 1)=p(1,1,0)=1/4$. It can be seen that observing both $X_{t-1}$ and $Y_{t-1}$ allows the future state $Y_t$ to be determined with certainty, while $X_{t-1}$ alone is not predictive of $Y_t$, since $X_{t-1}$ and $Y_t$ are statistically independent. The TE with memory parameter $m=n=1$ is given as $\textrm{TE}_{X\rightarrow Y}(1,1)=I(X_{t-1}; Y_t|Y_{t-1})=1$ bit, although there is no \emph{intrinsic} information flow between the two sequences but only a synergistic mechanism relating both $Y_{t-1}$ and $X_{t-1}$ to $Y_t$. 
\QEDA

In order to distinguish intrinsic and synergistic information flows, reference \cite{RBBJ:18} proposed to decompose the TE into Intrinsic Transfer Entropy (ITE) and Synergistic Transfer Entropy (STE). The ITE aims to capture the amount of information on $Y_t$ that is contained in the past of $\{X_t\}$ in addition to that already present in the past of $\{Y_t\}$; while the STE measures the information about $Y_t$ that is obtained only when combining the past of both $\{X_t\}$ and $\{Y_t\}$. Formally, the ITE from process $\{X_t\}$ to process $\{Y_t\}$ with memory parameters $(m,n)$ is defined as \cite{RBBJ:18}
\begin{align}   \label{def:ITE1}
\textrm{ITE}_{X\rightarrow Y}(m,n)& \defeq \underset{p(\bar{y}^{t-1}_{t-n}|y_{t-n}^{t-1})}{\mathrm{inf}}~I(X^{t-1}_{t-m};Y_{t}|\bar{Y}^{t-1}_{t-n}).
\end{align}
In definition \eqref{def:ITE1}, auxiliary variables $\bar{Y}^{t-1}_{t-n}$ can take values without loss of generality in the same alphabet as the corresponding variables $Y^{t-1}_{t-n}$ \cite{JD:03}, and are obtained by optimising the conditional distribution $p(\bar{y}^{t-1}_{t-n}|y_{t-n}^{t-1})$. The quantity \eqref{def:ITE1} can be shown to be an upper bound on the size (in bits) of a secret key that can be generated by two parties, one holding $X_{t-m}^{t-1}$ and the other $Y_t$, via public communication when the adversary has $Y_{t-n}^{t-1}$ \cite{US:99}. This intuitively justifies its use as a measure of intrinsic information flow. The STE is then defined as the residual 
\begin{align}\label{ste}
\textrm{STE}_{X\rightarrow Y}(m,n)\defeq\textrm{TE}_{X\rightarrow Y}(m,n)-\textrm{ITE}_{X\rightarrow Y}(m,n).
\end{align}


\subsection{TE and DI Estimation}
The TE can be estimated using tools akin to the estimation of MI, including plug-in methods \cite{Freedman1981}, non-parametric techniques based on kernel \cite{T:00} or  k-nearest-neighbor (k-NN) methods \cite{KSG:04,FP:07}, and parametric techniques, such as Maximum Likelihood \cite{SSSK:08} or Bayesian estimators \cite{WW:95}. Popular implementations of some of these standard methods can be found in the Java Information Dynamics Toolkit (JIDT) \cite{T:14} and TRENTOOL toolbox \cite{CMRVM:11}.
For the DI, estimators have been designed that rely on parametric and non-parametric techniques, making use also of universal compressors \cite{JHLYT:13, Quinn2011,MKTA:16}. In order to enable scaling over large data sets and/or data dimensions, MI estimators that leverage neural networks have been recently the subject of numerous studies. Notably, reference \cite{MASSYAD:18} introduced the Mutual Information Neural Estimator (MINE), which reduces the problem of estimating MI to that of classifying dependent vs. independent pairs of samples via the Donsker-Varadhan (DV) variational equality. Specifically, reference \cite{MASSYAD:18} proposes to train a neural network to approximate the solution of the optimization problem defined by the DV equality. The follow-up paper \cite{SHS:09} proposes to train a two-sample neural network classifier, which is then used as an approximation of the likelihood ratio in the DV equality. Theoretical limitations of general variational MI estimators were derived in \cite{JS:19}, which also proposes a variational MI estimator with reduced variance. We note that reference \cite{SHS:09} also considers the estimation of the conditional MI, which applies directly to the estimate of the TE as discussed in Section~\ref{back}. 

\subsection{Main Contributions, Paper Organization, and Notation}

This work proposes an estimator, referred to as ITE Neural Estimator (ITENE), of the ITE that is based on two-sample classifier and on the pathwise estimator of Monte Carlo gradients, also known as reparameterization trick \cite{MRFM:19}. We also present numerical results to illustrate the performance of the proposed estimator. The paper is organized as follows. In Section \ref{back}, we review the classifier-based MINE approach proposed in reference \cite{SHS:09}. Based on this approach, we introduce the proposed ITENE method in Section \ref{ITE}. Section \ref{experiments} presents experimental results. Throughout this paper, we use uppercase letters to denote random variables and corresponding lowercase letters to denote their realizations. $\log$ represents the natural logarithm. $\nabla_x f(x)$ represents the gradient of scalar function $f(x)$ and $\text{J}_x f(x)$ the Jacobian matrix of vector function $f(x)$.  


\section{Background: Classifier-based Mutual Information Neural Estimator (MINE)} \label{back}


In this section, we review the classifier-based MINE for the estimation of the MI $I(U;V)$ between jointly distributed continuous random variables $U$ and $V$.  The MI satisfies the DV variational representation \cite{MS:83}
\begin{subequations}\label{mine}
\begin{align} 
I(U;V)&=\sup_{f(u,v)}~\textrm{E}_{p(u,v)}[f(U,V)]-\log (\textrm{E}_{p(u)p(v)}[e^{f(U,V)}])\label{mine2} \\
&=\sup_{r(u,v)}~\textrm{E}_{p(u,v)}\bigg[\log \Big(\frac{r(U,V)}{\textrm{E}_{p(u)p(v)}[r(U,V)]}\Big)\bigg],\label{mine3}
\end{align}
\end{subequations}
where the supremum is taken over all functions $f(U,V)$ in \eqref{mine2} and $r(U,V)=e^{f(U,V)}$ in \eqref{mine3} such that the two expectations in \eqref{mine2} are finite. Note that \eqref{mine} contains expectations both over the joint distribution $p(u,v)$ of $U$ and $V$ and over the product of the marginals $p(u)$ and $p(v)$. Intuitively, the functions $f(u,v)$ and $r(u,v)$ act as classifiers of a sample $(u,v)$ being either generated by the joint distribution $p(u,v)$ or by the product distribution $p(u)p(v)$. This is done by functions $f(u,v)$ and $r(u,v)$ ideally outputing a larger value in the former case than in the latter \cite[Chapter 6]{O:17}. More precisely, following \cite{JS:19}, we can interpret function $r(u,v)$ as an unnormalized estimate of the likelihood ratio $p(u,v)/\big(p(u)p(v)\big)$, with $\tilde{r}(U,V)=r(U,V)/\textrm{E}_{p(u)p(v)}[r(U,V)]$ being its normalized version. This normalization ensures the condition $\textrm{E}_{p(u)p(v)}[\tilde{r}(U,V)]=1$, which is satisfied by the true likelihood ratio $p(u,v)/(p(u)p(v))$ \cite{JS:19}. Mathematically, the supremum in \eqref{mine3} is achieved when $r(u,v)$ is equal to the likelihood ratio \cite[Theorem 1]{JS:19}, i.e., 
\begin{align} \label{optimal}
r^*(u,v)=\frac{p(u,v)}{p(u)p(v)}.
\end{align} 

\begin{algorithm}[t!] 
\caption{Classifier Based MINE \cite{MASSYAD:18, SHS:09}}\label{Alg1:MINE}
\begin{algorithmic}[1]
\State{\textbf{Input:}} 
\Statex~~{$\mathcal{D}_1=\{(u_t,v_t)\}_{t=1}^{T}$: observed data samples} 
\State{\textbf{Output:}}
\Statex~~{ $\hat{I}(U;V)$: mutual information estimate}
\vspace{5pt}
\State{obtain data set $\mathcal{D}_0=\{(u_n, v_{\pi(n)})\}_{n=1}^{T}$, where $\pi(n)$ is sampled i.i.d. from set $\{1,\dots,T\}$}
\State{label samples $i\in\mathcal{D}_1$ as $a=1$ and $j\in\mathcal{D}_0$ as $a=0$ to create labeled data sets $\bar{\mathcal{D}}_1$ and $\bar{\mathcal{D}}_0$}
\State{$\theta \leftarrow$ initialize neural network parameters} 
\State{$\alpha \leftarrow$ set learning rate} 
\State{$\tau \leftarrow$ set hyperparameter} 
\State{split $\bar{\mathcal{D}}_1$ into two subsets $\bar{\mathcal{D}}_{1,t}$ (training) and $\bar{\mathcal{D}}_{1,e}$} (estimation)
\State{split $\bar{\mathcal{D}}_0$ into two subsets $\bar{\mathcal{D}}_{0,t}$ (training) and $\bar{\mathcal{D}}_{0,e}$} (estimation)
\State{train binary classifier using training set $\{\bar{\mathcal{D}}_{1,t}, \bar{\mathcal{D}}_{0,t}\}$}
\State{output: $\hat{I}(U;V)=\frac{1}{|\bar{\mathcal{D}}_{1,e}|}\sum_{i\in\bar{\mathcal{D}}_{1,e}}\log \frac{p_{\theta}(a=1|i)}{1-p_{\theta}(a=1|i)}-\log \big(\frac{1}{|\bar{\mathcal{D}}_{0,e}|}\sum_{j\in\bar{\mathcal{D}}_{0,e}}\text{clip}_{\tau}(\frac{p_{\theta}(a=1|j)}{1-p_{\theta}(a=1|j)})\big)$}
\end{algorithmic}
\end{algorithm}

This observation motivates the classifier-based estimator introduced in \cite{SHS:09}. To elaborate, given a data set $\mathcal{D}=\{(u_t,v_t)\}_{t=1}^{T}$ of $T$ data points from the joint distribution $p(u,v)$, we label the samples with a target value $a=1$. Furthermore, we construct a data set $\mathcal{D}_0$ approximately distributed according to the product distribution $p(u)p(v)$ by randomly resampling the values of $v_t$ (see line 3 in Algorithm 1). These samples are labeled as $a=0$. We use notation $p(a=1|u,v)$ to represent the posterior probability that a sample is generated from the distribution $p(u,v)$ when the hypotheses $a=1$ and $a=0$ are a priori equally likely. An estimate of the probability $p(a=1|u,v)$ can be obtained by training a function $p_{\theta}(a=1|u,v)$ parametrized as a neural network with input $u$ and $v$, target output $a$, and weight vector $\theta$. This is done via the minimization of the empirical cross-entropy loss  evaluated on the described data sets (see lines 8-10 in Algorithm 1) via Stochastic Gradient Descent (SGD) (see, e.g., \cite[Chapter 6]{O:17}). Having completed training, the likelihood ratio can be estimated as 
\begin{align}  \label{rclip}
\hat{r}_{\theta}(u,v)=\frac{p_{\theta}(a=1|u,v)}{1-p_{\theta}(a=1|u,v)}. 
\end{align}
This follows since, at convergence, if training is successful, the following equality holds approximately
\begin{align}
p_{\theta}(a=1|u,v) = \frac{p(a=1)p(u,v|a=1)}{p(a=1)p(u,v|a=1)+p(a=0)p(u,v|a=0)} =\frac{ p(u,v)}{p(u,v)+p(u)p(v)}.
\end{align}
Finally, the estimate \eqref{rclip} can be plugged into an empirical approximation of \eqref{mine3} as
\begin{align}  \label{MINE}
\hat{I}(U;V)=\textrm{E}_{\hat{p}(u,v)} \Big[\log\Big(\frac{\hat{r}_{\theta}(U,V)}{\textrm{E}_{\hat{p}(u)\hat{p}(v)}[
\text{clip}_{\tau}(\hat{r}_{\theta}(U,V))]}\Big)\Big], 
\end{align}
where  $\hat{p}(u,v)$ represents the empirical distribution of the observed data sample pairs in an held-out part of data set $\mathcal{D}_1$, while $\hat{p}(u)$ and $\hat{p}(v)$ are the corresponding empirical marginal distributions for $U$ and $V$ (see line 11 in Algorithm 1); and the clip function is defined as $\text{clip}_{\tau}(v)=\max\{\min\{v,e^{\tau}\},e^{-\tau}\}$ with some constant $\tau\geq 0$ \cite{JS:19}. Clipping was suggested in \cite{JS:19} in order to reduce variance of the estimate \eqref{MINE}, and a similar approach is also used in \cite{SHS:09}. The estimator \eqref{MINE} is known to be consistent but biased \cite{MASSYAD:18}, and an analysis of the variance can be found in \cite{JS:19} (see also Lemma 1 below). Details are presented in Algorithm 1.

\section{Intrinsic Transfer Entropy Neural Estimator (ITENE)} \label{ITE}

In this section, inspired by the classifier-based MINE, we introduce an estimator for the ITE, which we refer to as ITENE. Throughout this section, we assume the availability of data in the form of time series $\mathcal{D}=\{(x_t, y_t): t=1,2,\dots,T\}$ generated as a realization of jointly stationary random processes $\{X_t, Y_t\}_{t\geq 1}$. We use the notations $X^-_t \defeq X_{t-m}^{t-1}$, $Y^-_t\defeq Y_{t-n}^{t-1}$ and $Y^0_t \defeq Y_t$ and we also drop the subscript $t$ when no confusion may arise.

\subsection{TENE}
We start by noting that, using the chain rule \cite{CT:06}, the TE in \eqref{def:te} can be written as the difference
\begin{align} \label{te}
\textrm{TE}_{X\rightarrow Y}(m,n)=I(X^{-}; Y^0, Y^{-})-I(X^{-}; Y^{-}).
\end{align} 
Therefore, the TE can be estimated by applying the classifier-based MINE in Algorithm 1 to both terms in \eqref{te} separately. This approach was proposed in \cite{SHS:09} and found empirically to outperform other estimates of the conditional MI. Accordingly, we have the estimate
\begin{align}\label{tente}
\widehat{\textrm{TE}}_{X\rightarrow Y}(m,n)=\hat{I}(X^-;Y^0, Y^-) - \hat{I}(X^-;Y^-),
\end{align}
where the MINE estimates in \eqref{MINE} are obtained by applying Algorithm 1 to the data sets $\mathcal{D}^A_1=\{u_t=x^-_t,v_t=(y^0_t, y^-_t)\}_{t=1}^{T}$ and $\mathcal{D}^B_1=\{u_t=x^-_t, v_t=y^-_t\}_{t=1}^{T}$, respectively (zero padding is used for out-of-range indices). We refer to the resulting estimator \eqref{tente} as TENE. Following \cite{SHS:09}, TENE is consistent but biased. Furthermore, without using clipping, i.e., when $\tau \rightarrow \infty$, we have that the following lemma holds. 
\begin{lemma}
Assume that the estimates $\hat{r}_{\theta}(x^-, y^0, y^-)$ and $\hat{r}_{\theta}(x^-, y^-)$ equal their respective true likelihood ratios, i.e., $\hat{r}_{\theta}(x^-, y^0, y^-)=p(x^-, y^0, y^-)/(p(x^-)p(y^0, y^-))$ and $\hat{r}_{\theta}(x^-,y^-)=p(x^-, y^-)/(p(y^-)p(y^-))$. Then, under the randomness of the sampling procedure generating the data set $\mathcal{D}$, we have 
\begin{align}
\lim_{T\rightarrow \infty}T~\text{Var}[\widehat{\textrm{TE}}_{X\rightarrow Y}(m,n)] \geq e^{I(X^{-}; Y^0, Y^{-})}+e^{I(X^{-};Y^{-})}-2.
\end{align}
\end{lemma}

The proof follows directly from \cite[Theorem 1]{JS:19}. Lemma 1 demonstrates that, without clipping, the variance of TENE in \eqref{tente} can grow exponentially with the maximum of the true values of $I(X^{-}; Y^0, Y^{-})$ and $I(X^{-}; Y^{-})$. Note that a similar result applies to MINE \cite{JS:19}. Setting a suitable value for $\tau$ is hence important in order to obtain reliable estimates. 
\subsection{ITENE}
We now move on to the estimator of the ITE \eqref{def:ITE1}. To this end, we first parameterize the distribution $p_{\phi}(\bar{y}^-|y^-)$ under optimization as 

\begin{align} \label{transform}
\bar{y}^-_{\phi}=\mu_{\phi}(y^-)+\sigma_{\phi}(y^-) \odot \epsilon,
\end{align} 
where $\mu_{\phi}(y^-)$ and $\log\sigma_{\phi}(y^-)$ are disjoint sets of outputs of a neural network with weights $\phi$; $\odot$ is the element-wise product; and $\epsilon \sim \mathcal{N} (0,\textbf{I})$ is a Gaussian vector independent of all other variables. Parameterization \eqref{transform} follows the so-called reparameterization trick popularized by the variational auto-encoder \cite{KW:13}. An estimator of the ITE \eqref{def:ITE1} can be defined by optimizing over $\phi$ the ITE \eqref{te} as 
\begin{align}\label{estimation}
\widehat{\textrm{ITE}}_{X\rightarrow Y}(m,n)=\underset{\phi}{\mathrm{inf}} \big(\hat{I}_{\phi}(X^-;Y^0,\bar{Y}^-)-\hat{I}_{\phi}(X^-;\bar{Y}^-)\big),
\end{align}
where we have made explicit the dependence of estimates $\hat{I}_{\phi}(X^-;Y^0,\bar{Y}^-)$ and $\hat{I}_{\phi}(X^-;\bar{Y}^-)$ on $\phi$. In particular, using \eqref{te}, the first MINE estimate in \eqref{tente} can be written as a function of $\phi$ as
\begin{align} \label{Areform}
\hat{I}_{\phi}(X^-;Y^0,\bar{Y}^-)=\textrm{E}_{\hat{p}(x^-,y^0,y^-)}\big [\mathrm{E}_{p(\epsilon)} [\log(\hat{r}_{\theta}(X^-,Y^0,\bar{Y}^-_{\phi}))]\big]-\log(\textrm{E}_{\hat{p}(x^-)\hat{p}(y^0,y^-)}\big[\mathrm{E}_{p(\epsilon)}[\text{clip}_{\tau}(\hat{r}_{\theta}(X^-,Y^0,\bar{Y}^-_{\phi}))]\big], 
\end{align}
where parameter $\theta$ is obtained from Algorithm 1 by considering as input the data set $\mathcal{D}_{\phi,1}^{A}=\{u_t = x^-_t, v_t =( y_t^0, \bar{y}_{\phi,t}^-)\}_{t=1}^{T}$, where samples $\bar{y}_{\phi,t}^-$ are generated using \eqref{transform} as $\bar{y}^-_{\phi,t}=\mu_\phi(\bar{y}_t)+\sigma_{\phi}(\bar{y}_t) \odot \epsilon_t$ for i.i.d. samples $\epsilon_t \sim \mathcal{N} (0,\textbf{I})$. Furthermore, the empirical distributions $\hat{p}(\cdot)$ in \eqref{Areform} are obtained from the held-out (estimation) data set in Algorithm 1. In a similar manner, the second MINE estimate in \eqref{estimation} is given as 
\begin{align} \label{Breform} 
\hat{I}_{\phi}(X^-;\bar{Y}^-)=\textrm{E}_{\hat{p}(x^-,y^-)}\big [\mathrm{E}_{p(\epsilon)} [\log(\hat{r}_{\theta'}(X^-,\bar{Y}^-_{\phi}))]\big]-\log(\textrm{E}_{\hat{p}(x^-)\hat{p}(y^-)}\big[\mathrm{E}_{p(\epsilon)}[\text{clip}_{\tau}(\hat{r}_{\theta'}(X^-,\bar{Y}^-_{\phi}))]\big], 
\end{align}
where parameter $\theta'$ is obtained from Algorithm 1 by considering as input the data set $\mathcal{D}_{\phi,1}^B=\{u_t = x^-_t, v_t =\bar{y}_{\phi,t}^-)\}_{t=1}^{T}$.


\begin{algorithm}[t!] 
\caption{ITENE}\label{Alg1}
\begin{algorithmic}[1]
\State{\textbf{Input:}} 
\Statex~~{$\mathcal{D}=\{(x_t, y_t)\}_{t=1}^{T}$: observed data samples from the random process $\{X_t, Y_t\}$} 

\State{\textbf{Output:}}
\vspace{3pt}
\Statex~~{ $\widehat{\textrm{ITE}}_{X\rightarrow Y}(m,n)$: ITE estimate}  
\vspace{8pt}
\State{$(\phi,\theta,\theta')\leftarrow$ initialize network parameters}
\State{$\alpha\leftarrow$ set learning rate}
\State{$\tau \leftarrow$ set hyperparameter}  

\State{\textbf{repeat}}
\State~~{randomly generate $T$ samples $\{\epsilon_t\}_{t=1}^{T}$ from distribution $ \mathcal{N} (0,\textbf{I})$}
\State~~{for each $t=1,\dots,T$:}
\State~~~~~~{compute $\bar{y}_{\phi,t}^-=\mu_{\phi}(y_t^-)+\sigma_{\phi}(y_t^-) \odot \epsilon_t$}
\State~~{define data set $\mathcal{D}^A=\{u^A_t,v^A_t\}_{t=1}^{T}$, with $u^A_t=x^-_t, v^A_t=\{y^0_t,\bar{y}^-_{\phi,t}\}$}
\State~~{apply Algorithm 1 to output $\hat{I}_{\phi}(X^-;Y^0,Y^-)=\hat{I}(U^{A};V^{A})$}
\State~~{define data set $\mathcal{D}^B=\{u^B_t,v^B_t\}_{t=1}^{T}$, with $u^B_t=x^-_t, v^B_t=\bar{y}^-_{\phi,t}$}
\State~~{apply Algorithm 1 to output $\hat{I}_{\phi}(X^-;Y^-)=\hat{I}(U^{B};V^{B})$}
\State~~{update the network parameters using the pathwise gradient estimators \eqref{gra:A}-\eqref{gra:B}}
\State~~~~~~{$\phi\leftarrow \phi-\alpha\nabla_{\phi}\big(\hat{I}_{\phi}(X^-;Y^0,Y^-)-\hat{I}_{\phi}(X^-;Y^-)\big)$}
\State{\textbf{until} convergence}
\State{output: $\widehat{\textrm{ITE}}_{X\rightarrow Y}(m,n)=\hat{I}_{\phi}(X^-;Y^0,Y^-)-\hat{I}_{\phi}(X^-;Y^-)$}
\end{algorithmic}
\end{algorithm}

We propose to tackle problem \eqref{estimation} in a block coordinate fashion by iterating between SGD steps with respect to $\phi$ and updates of parameters $(\theta,\theta')$ using Algorithm 1. To this end, when fixing $(\theta,\theta')$, the optimization over parameter $\phi$ requires the gradient
\begin{align} 
\nabla_{\phi}\hat{I}_{\phi}(X^-;Y^0,\bar{Y}^-)=\mathrm{E}_{\hat{p}(x^-,y^0,y^-)}\Bigg[\mathrm{E}_{p(\epsilon)}\Bigg[\frac{\nabla_{\bar{y}^-_\phi}\hat{r}_{\theta}}{\hat{r}_{\theta} } \times  \text{J}_{\phi}\bar{y}^-_\phi \Bigg]\Bigg]-\frac{\mathrm{E}_{\hat{p}(x^-)\hat{p}(y^0,y^-)}\big[\mathrm{E}_{p(\epsilon)}[\nabla_{\bar{y}^-_\phi}\hat{r}_{\theta} \times  \text{J}_{\phi}\bar{y}^-_\phi]\big]}{\mathrm{E}_{\hat{p}(x^-)\hat{p}(y^0,y^-)}\big[\mathrm{E}_{p(\epsilon)}[\hat{r}_{\theta}]\big]} \label{gra:A},  
\end{align}
where, from \eqref{rclip}, we have the gradient  
\begin{align}
\nabla_{\bar{y}^-_\phi}\hat{r}_{\theta}=\frac{\nabla_{\bar{y}^-_\phi}p_{\theta}(a=1|x^0,y^-,\bar{y}^-_\phi)}{(1-p_{\theta}(a=1|x^0,y^-,\bar{y}^-_\phi))^2};
\end{align}
and, from \eqref{transform}, we have the Jacobian $\text{J}_{\phi}\bar{y}^-_\phi=\text{J}_{\phi}\mu_{\phi}(Y^-)+\big(\text{J}_{\phi}(\sigma_{\phi}(Y^-)\big) \odot \epsilon$. It also requires the gradient 
\begin{align}
\nabla_{\phi}\hat{I}_{\phi}(X^-;\bar{Y}^-)=\mathrm{E}_{\hat{p}(x^-,y^-)}\Bigg[\mathrm{E}_{p(\epsilon)}\Bigg[\frac{\nabla_{\bar{y}^-_{\phi}}\hat{r}_{\theta'}}{\hat{r}_{\theta'}} \times  \text{J}_{\phi}\bar{y}^-_{\phi}\Bigg]\Bigg]-\frac{\mathrm{E}_{\hat{p}(x^-)\hat{p}(y^-)}\big[\mathrm{E}_{p(\epsilon)}[\nabla_{\bar{y}^-_\phi}\hat{r}_{\theta'} \times  \text{J}_{\phi}\bar{y}^-_{\phi}]\big]}{\mathrm{E}_{\hat{p}(x^-)\hat{p}(y^-)}\big[\mathrm{E}_{p(\epsilon)}[\hat{r}_{\theta'}]\big]}  \label{gra:B},
\end{align}
where we have 
\begin{align}
\nabla_{\bar{y}^-_{\phi}}\hat{r}_{\theta'}=\frac{\nabla_{\bar{y}^-_\phi}p_{\theta'}(a=1|x^0,\bar{y}^-_\phi)}{(1-p_{\theta'}(a=1|x^0,\bar{y}^-_\phi))^2}.
\end{align}
We note that the gradients \eqref{gra:A}-\eqref{gra:B} are instances of pathwise gradient estimators \cite{MRFM:19}. The resulting ITENE is summarized in Algorithm~\ref{Alg1}. Due to the consistency of TENE, ITENE is also consistent if the capacity of the model $p_{\phi}$ is large enough.

\section{Experiments} \label{experiments}
In this section, we provide some results to illustrate the type of insights that can be obtained by decomposing the TE into ITE and STE as in \eqref{ste}. To this end, consider first the following simple example. The joint processes $\{X_t, Y_t\}_{t\geq 1}$ are generated according to 
\begin{align} \label{examp}
Y_t=\left\{
  \begin{array}{ll}
	Z_t, &\text{if}~Y_{t-1} <\lambda  \\
	\rho X_{t-1}+\sqrt{1-\rho^2} Z_t, &\text{if}~Y_{t-1} \geq \lambda,
	\end{array}
		\right.
\end{align}
for some threshold $\lambda$, where variables $\{X_t, Y_t\}$ are independent and identically distributed as $\mathcal{N} (0,1)$. Intuitively, for large values of the threshold $\lambda$, there is no information flow between $\{X_t\}$ and $\{Y_t\}$, while for small values, there is a purely intrinsic flow of information. For intermediate values of $\lambda$, the information flow is partly synergistic, since knowing both $Y_{t-1}$ and $X_{t-1}$ is instrumental in obtaining information about $Y_t$. To quantify the intuition above, we apply the discussed estimators with $m=n=1$. To this end, for all two-sample neural network classifiers, we consider two hidden layers with 100 hidden neurons with ELU activation functions, while for the probability $p_{\phi}(\bar{y}^-|y^-)$, we adopt a neural network with hidden layer of 200 neurons with ELU activation functions and outputs $\mu_{\phi}(y^-)$ and $\log(\sigma_{\phi}(y^-))$. The data set size $T$ is split into a $75\%$-fraction for classifier training and a $25\%$-fraction for estimation. We set learning rate $\alpha=0.001$ and clipping parameter $\tau=0.9$.

\begin{figure}[t!] 
  \centering
\includegraphics[width=0.45\columnwidth]{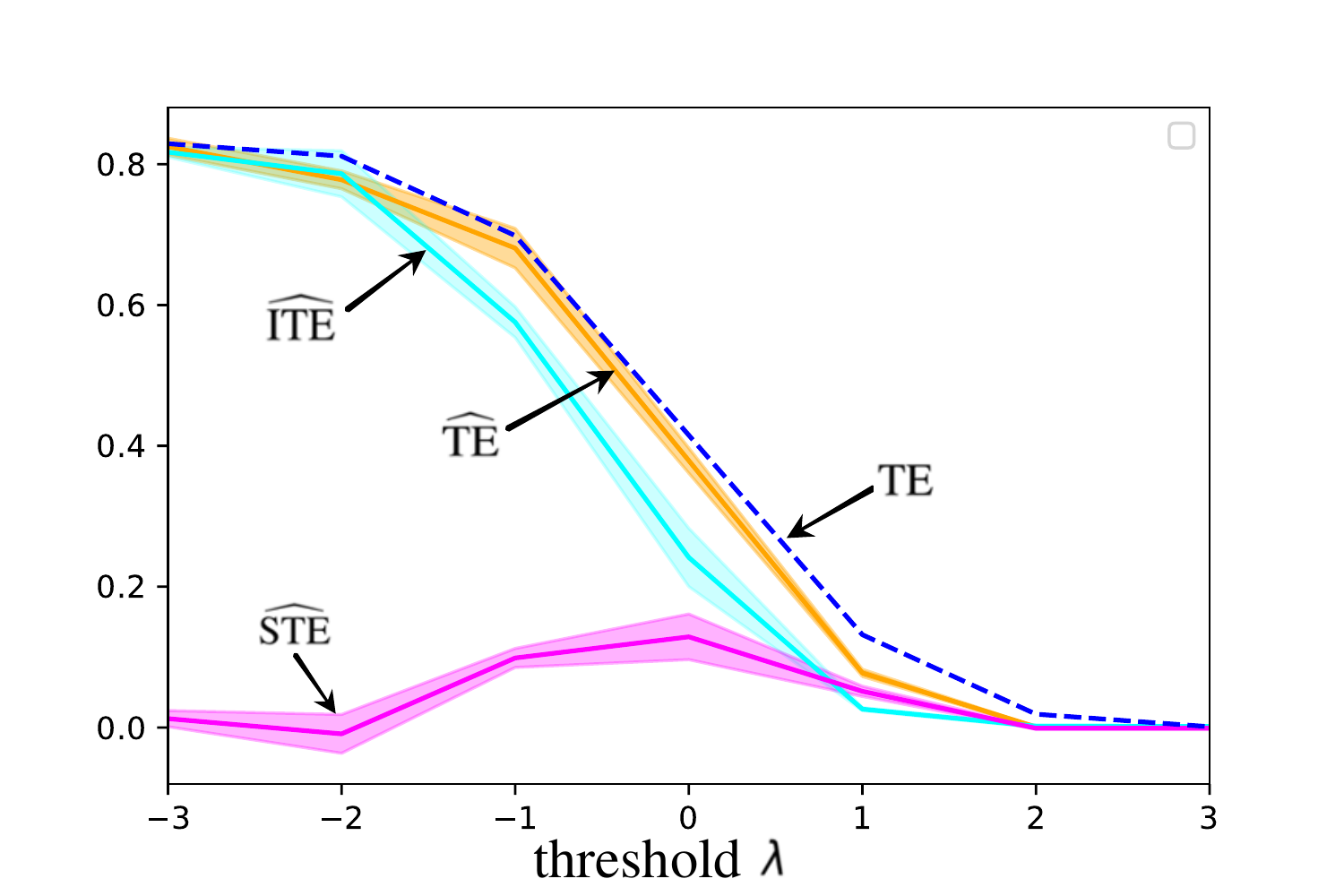}
\caption{TENE, ITENE,  STENE (obtained as the difference \eqref{ste}) and true TE versus threshold $\lambda$ with $\rho=0.9$ for the example \eqref{examp}. Dashed areas represent the range of observed estimates within 10 trials.}
\label{fig:examplep}
\end{figure}

\begin{figure}[t!] 
  \centering
\includegraphics[width=0.43\columnwidth]{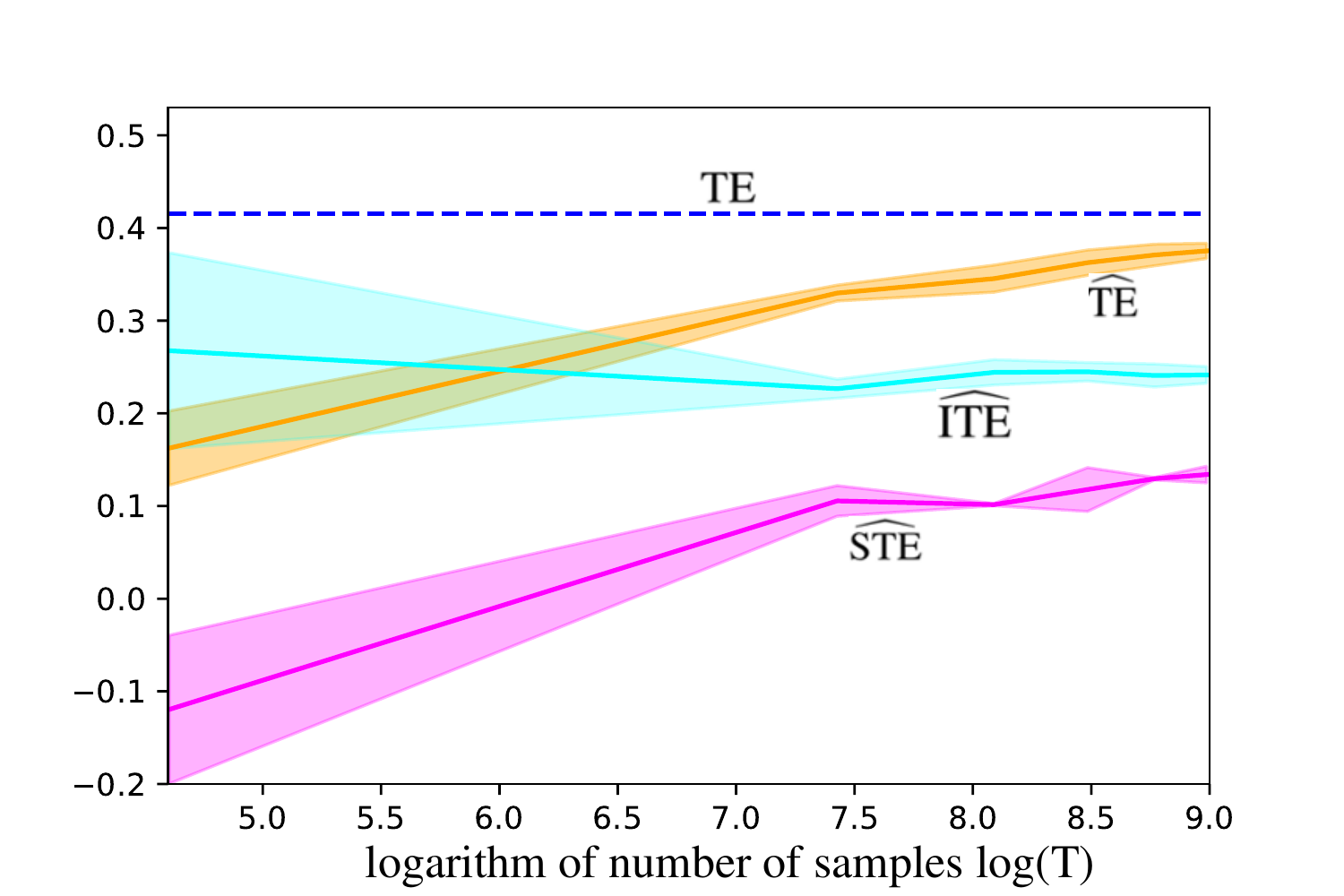}
\caption{TENE, ITENE, STENE (obtained as the difference \eqref{ste}) and true TE versus the number $T$ of samples with $\rho=0.9$ and $\lambda=0$ for the example \eqref{examp}. Dashed areas represent the range of observed estimates within 10 trials.}
\label{fig:example}
\end{figure}

The computed estimates $\widehat{\textrm{TE}}_{X\rightarrow Y}(1,1)$, $\widehat{\textrm{ITE}}_{X\rightarrow Y}(1,1)$, $\widehat{\textrm{STE}}_{X\rightarrow Y}(1,1)$ are plotted in Fig.~\ref{fig:examplep} and Fig.~\ref{fig:example} as a function of the threshold $\lambda$ and the number of samples $T$, respectively, along with the true TE. The latter can be computed in closed form as $\textrm{TE}_{X\rightarrow Y}(m,n)=\textrm{TE}_{X\rightarrow Y}(1,1)=-0.5Q(\lambda)\log(1-\rho^2)$ (nats), where $Q(\cdot)$ is the standard complementary cumulative distribution function of a standard Gaussian variable. In a manner consistent with the intuition provided above, when $\lambda$ is either small, i.e., $\lambda\leq -2$, or large, i.e., $\lambda\geq 2$,  the ITE is seen in Fig.~\ref{fig:examplep} to be close to the TE, yielding nearly zero STE. This is not the case for intermediate values of $\lambda$, in which regime a non-negligible STE is observed. 

In Fig.~\ref{fig:example}, we investigate the impact of the number $T$ of samples when $\lambda=0$, at which point the gap between the ITE and the TE is the largest (see Fig.~\ref{fig:examplep}). As illustrated in the figure, the four estimates becomes increasingly accurate as $T$ increases, reflecting the consistency of the estimators.  

\begin{figure}[t!] 
  \centering
\includegraphics[width=0.45\columnwidth]{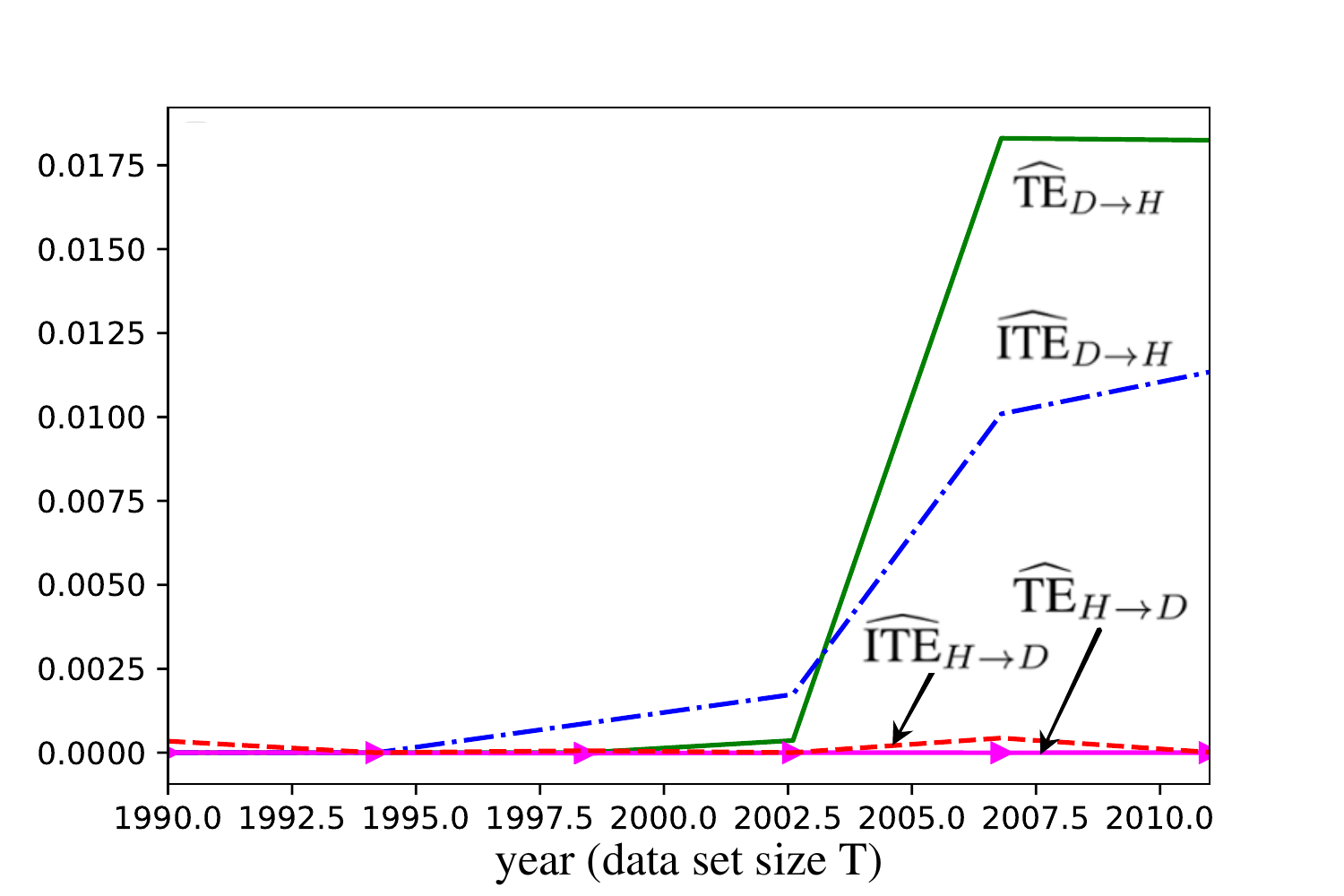}
\caption{TENE and ITENE between the DJIA, denoted as "D", and the HSI, denoted as "H".}
\label{fig:stock}
\end{figure}

For a real-world example, we apply the estimators at hand to historic data of the values of the Hang Seng Index (HSI) and of the Dow Jones Index (DJIA) between 1990 and 2011.  As done in \cite{JHLYT:13}, for each stock, we classify its values into three levels, namely $1$, $0$, and $-1$, where $1$ indicates an increase in the stock price by more than $0.8\% $ in one day, $-1$ indicates a drop by more than $-0.8\%$, and $0$ indicates all other cases. As illustrated in Fig.~\ref{fig:stock}, and in line with the results in \cite{JHLYT:13}, both the TE and ITE from the DJIA to the HSI are much larger than in the reverse direction, implying that the DJIA influenced the HSI more significantly than the other way around for the given time range. Furthermore, we observe that not all the information flow is estimated to be intrinsic, and hence the joint observation of the history of the DJIA and of the HSI is partly responsible for the predictability of the HSI from the DJIA. 


\section{Conclusions} \label{conclusions}
In this work, we have proposed an estimator for Intrinsic Transfer Entropy (ITE) between two time series based on two-sample neural network classifiers and the reparameterization trick. As future work, it would be interesting to apply the estimator to larger-scale data sets and to further investigate its theoretical properties. 

\section*{Acknowledgements}
Jingjing Zhang and Osvaldo Simeone have received funding from the European Research Council (ERC) under the European Union's Horizon 2020 Research and Innovation Programme (Grant Agreement No. 725731). Jingjing Zhang has also been supported by a King's Together award. 

\bibliographystyle{IEEEtran}
\bibliography{IEEEabrv,final_refs}

\end{document}